\documentclass[%
reprint,
amsmath,amssymb,
aps,
]{revtex4-1}

\usepackage[dvipdfmx]{graphicx}
\usepackage{physics}
\usepackage{graphicx}% Include figure files
\usepackage{dcolumn}% Align table columns on decimal point
\usepackage{bm}% bold math
\usepackage{color,ulem}
\usepackage{braket}
\usepackage{mathrsfs}
\usepackage{amsmath}
\usepackage{amssymb}
\usepackage{float}
\usepackage{appendix}
\usepackage{comment}
\usepackage{ulem}
\newcommand{\mtx}[1]{\begin{pmatrix} #1 \end{pmatrix}}
\newcommand{\mysout}[1]{} %submitver

\begin{document}

% \preprint{}

\title{
Spacetime-Localized Response in Quantum Critical Spin Systems: Insights from Holography
}

\author{Motoaki Bamba${}^{1,2,3}$}
\author{Koji Hashimoto${}^3$}
\author{Keiju Murata${}^4$}
\author{Daichi Takeda${}^3$}
\author{Daisuke Yamamoto${}^4$}

\affiliation{${}^1$Department of Physics, Graduate School of Engineering Science, Yokohama National University, Yokohama 240-8501, Japan}
\affiliation{${}^2$The Hakubi Center for Advanced Research, Kyoto University, Kyoto 606-8501, Japan}
\affiliation{${}^3$Department of Physics, Kyoto University, Kyoto 606-8502, Japan}
\affiliation{${}^4$Department of Physics, College of Humanities and Sciences, Nihon University, Sakurajosui, Tokyo 156-8550, Japan}

\begin{abstract}
According to the AdS/CFT correspondence, certain quantum many-body systems in $d$-dimensions are equivalent to gravitational theories in $(d+1)$-dimensional asymptotically AdS spacetimes. When a massless particle is sent from the AdS boundary to the bulk curved spacetime, it reaches another point of the boundary after a time lag. In the dual quantum system, it should appear as if quasiparticles have been transferred between two separated points. We theoretically demonstrate that this phenomenon, which we call ``spacetime-localized response,'' is actually observed in the dynamics of the one-dimensional transverse-field Ising model near the quantum critical point. This result suggests that, if we can realize a holographic spin system in a laboratory, the experimental probing of the emergent extra-dimension is possible by applying a designed stimulus to a quantum many-body system, which is holographically equivalent to sending a massless particle through the higher-dimensional curved bulk geometry. We also discuss possible experimental realizations using Rydberg atoms in an optical tweezers array.
\end{abstract}

%\begin{description}
%\item[Usage]
%Secondary publications and information retrieval purposes.
%\item[Structure]
%You may use the \texttt{description} environment to structure your %abstract;
%use the optional argument of the \verb+\item+ command to give the %category of each item. 
%\end{description}

%\keywords{Suggested keywords}%Use show keys class option if keyword
%display desired
\maketitle

%\tableofcontents

%%%%%%%%%%%%%%%%%%%%%%%%%%%%%%%%%%%%%%%%%%%%%%%%%%%%%%%%%%%%%%%%%%%%%%%%%%%%%%%%%%%%%%%%%%%%%%%%%%%%%%%%%%%%%%%
\setcounter{footnote}{0}
\section{Introduction} 
The AdS/CFT correspondence~\cite{Maldacena:1997re,Gubser:1998bc,Witten:1998qj} is a holographic duality between the gravitational theory in the $(d+1)$-dimensional anti-de Sitter spacetime (AdS) and the conformal field theory (CFT) living in the $d$-dimensional boundary of the AdS. Although the correspondence is originally proposed for a supersymmetric large-$N$ Yang-Mills theory, its idea has been also applied to more realistic systems such as condensed matter systems~\cite{Hartnoll:2009sz,Herzog:2009xv,McGreevy:2009xe,Horowitz:2010gk,Sachdev:2010ch}.
Those studies indicate that there may be ``materials'' having their gravitational duals in our world.
Experimenting with such materials, if realized, opens a new path for tabletop experiments of quantum gravity.

\begin{figure}
\centering
\includegraphics[scale=0.27]{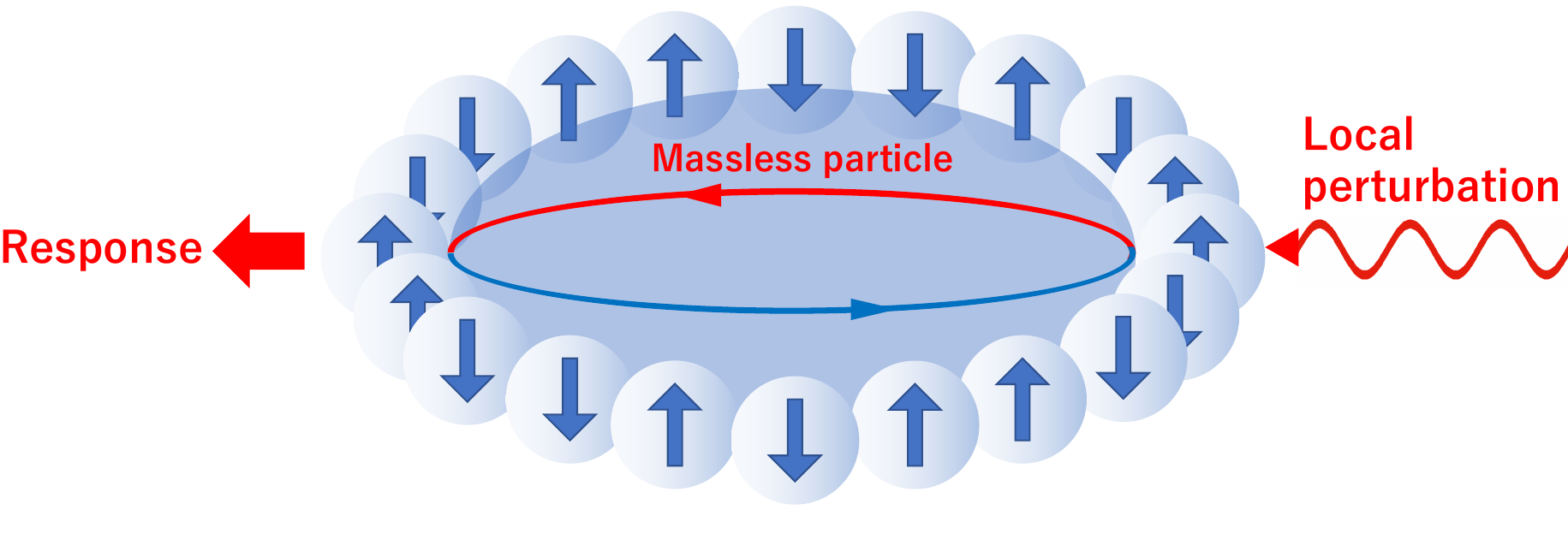}
\caption{Schematic picture of the spacetime-localized response in a spin system on a lattice ring. The response to a local perturbation suddenly appears at a spatially separated point after a certain time lag.
}
\label{ponchi}
\end{figure}

As a tool of probing the dual spacetime, some of the authors of this paper and others have proposed a way to create a null geodesic in the asymptotically AdS spacetime by the manipulation of the source in the quantum field theory (QFT)~\cite{Kinoshita:2023hgc,Hashimoto:2019jmw,Hashimoto:2018okj,Kaku:2021xqp,Hashimoto:2022aso, Terashima:2020uqu, Terashima:2021klf, Terashima:2023mcr}. 
Once a null geodesic is created in AdS, it bounces repeatedly at the AdS boundary~\cite{Kinoshita:2023hgc,Terashima:2023mcr}. 
In the viewpoint of the dual QFT, while the energy flux locally propagates following the conservation law, the operator expectation value coupled to the source has sharp peaks at spacetime-points where the null geodesic collides with the boundary.
We refer to this phenomenon as ``spacetime-localized response.''
The spacetime-localized response is naturally understood with the knowledge of the dual spacetime as schematically drawn in Fig.\ref{ponchi}, but is highly non-trivial in terms of the QFT.

\begin{figure}
\centering
\includegraphics[scale=0.5]{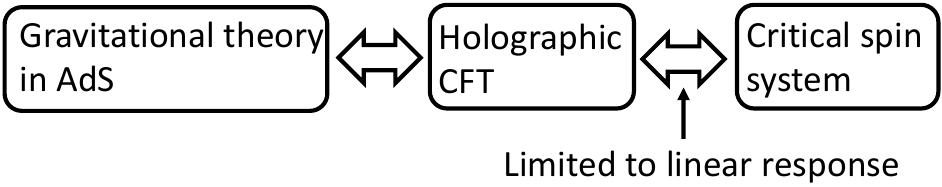}
\caption{In the linear response regime, geometrical interpretation is possible for (non-holographic) critical spin systems.
}
\label{theories}
\end{figure}

In this paper, we investigate the occurrence of the spacetime-localized response in the transverse-field Ising model on a lattice ring, aiming at its experimental realization. This model is known to be described by a CFT at the critical point. 
While it may not have a gravitational dual due to its small central charge, we demonstrate that the Jordan-Wigner fermions, generated by a specific form of local perturbation, do exhibit the spacetime-localized response. 
This should be attributed to the fact that the retarded propagators in CFTs are determined only from conformal dimensions of operators~\cite{Son:2002sd, Terashima:2020uqu, Terashima:2021klf, Terashima:2023mcr}. Because of the universality of the linear response in CFTs, we can still apply the ``geometrical'' interpretation inspired by the AdS/CFT even for some non-holographic spin systems. (See Fig.~\ref{theories}). This fact helps our intuitive understanding of non-trivial phenomena in spin systems.
Furthermore, we observe that the spacetime-localized response persists, even in small lattice systems, well outside the QFT limit. These observations suggest the universality of the spacetime-localized response for general quantum many-body models that possess an (approximate) CFT description. 

We also propose experimental realizations using a well-controllable quantum system of Rydberg atoms in an optical tweezers array~\cite{Browaeys:2020,Bernien:2017,Nogrette:2014,Ebadi:2021,Scholl:2021} in the case of the transverse-field Ising model.
Although our system may not be holographic, the technique developed in this paper, both theoretically and experimentally, can directly be applied to spin systems known to have large central charges, which will be possibly holographic.
If we can realize such holographic spin systems in a laboratory, we can probe dual spacetimes (e.g. black hole spacetime) by table-top experiments.
The observation of the spacetime-localized response in a laboratory setting can serve as a foundation for future experiments with holographic spin systems.
This study not only predicts the nontrivial phenomenon of spacetime-localized response, which is interesting in its own right, but also offers an experimental method to see the motion of particles in the dual higher-dimensional spacetime of a holographic spin system. 
Thus, experimenting with our setup represents a crucial first step in exploring semi-classical gravitational duals for condensed matter systems in tabletop experiments.

%%%%%%%%%%%%%%%%%%%%%%%%%%%%%%%%%%%%%%%%%%%%%%%%%%%%%%%%%%%%%%%%%%%%%%%%%%%%%%%%%%%%%%%%%%%%%%%%%%%%%%%%%%%%%%%
\begin{figure}
\centering
\includegraphics[scale=0.20]{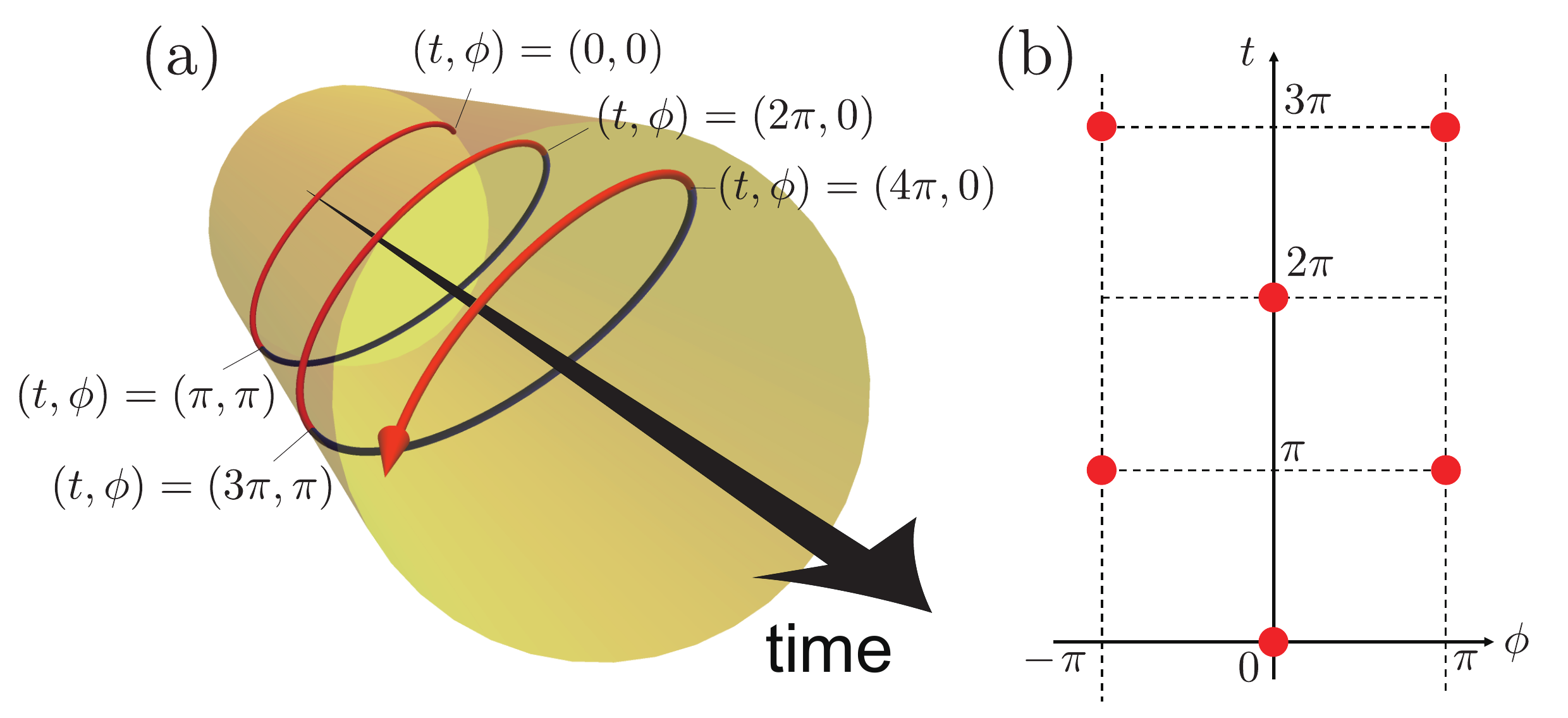}
\caption{(a) A typical orbit of the null geodesic in AdS$_3$. It bounces repeatedly at the AdS boundary depicted as the yellow cylinder. The radial distance $\rho\equiv r/\sqrt{1+r^2}$ and  the azimuthal angle $\phi$ correspond to space coordinates, while the axial coordinate corresponds to the time variable $t$. The AdS boundary is located at $\rho=1$ ($r\rightarrow \infty$). (b) Points at which the null geodesic collides with the AdS boundary on the $(t,\phi)$-plane. 
\label{nullg}
}
\end{figure}

\section{Null geodesics in AdS$_3$}
We consider the null geodesic in the global AdS$_3$ spacetime: 
\begin{equation}
ds^2=-(1+r^2)dt^2+\frac{dr^2}{1+r^2}+r^2d\phi^2\ ,
\end{equation}
where we take the unit of AdS radius $=1$. We have the exact solution of the null geodesic equation in the AdS$_3$ as 
\begin{equation}
 t=\frac{\pi}{2}+\tan^{-1} \lambda\ ,\  r^2=\frac{m^2+\lambda^2}{1-m^2}\ ,\ \phi=\frac{\pi}{2}+\tan^{-1}\frac{\lambda}{m}\ ,
\end{equation}
where $\lambda$ is the affine parameter and $m$ is the angular momentum per unit energy. % as $m$. %=M/\Omega$. 
A typical orbit of the null geodesic is shown in Fig.~\ref{nullg}{(a)}. 
Suppose that a null geodesic is injected into the AdS$_3$ from the AdS boundary at $(t,\phi)=(0,0)$. 
Such a particle arrives at the antipodal point $(t,\phi)=(\pi,\pi)$, bounces back there, and returns to the original position $(t,\phi)=(2\pi,0)$. 
Figure \ref{nullg}{(b)} shows the points at which the particle reaches the AdS boundary on $(t,\phi)$-plane.
These points are independent of the angular momentum $M$ of the null geodesic, although the trajectory does depend on it.

If a source is applied appropriately in the CFT living on the AdS boundary, a null geodesic can be produced in the bulk as a localized configuration of the probe field~\cite{Kinoshita:2023hgc}. 
The null geodesic with energy $\Omega$ and angular momentum $M$ is created by the source, 
\begin{equation}
 \mathcal{J}(t,\phi)=A\exp\left[
-\frac{t^2}{2\sigma_t^2}-\frac{\phi^2}{2\sigma_\phi^2}-i\Omega t +iM\phi
\right]\ ,
\label{src}
\end{equation}
where $A$, $\sigma_t$, and $\sigma_\phi$ are the amplitude and the widths in $t$ and $\phi$, respectively, of the Gaussian part. This source modifies the action of the CFT as $S\to S + \int dtd\phi\, \mathcal{J}\mathcal{O}$ where $\mathcal{O}$ is an operator in the CFT. 
We will consider the source with finite width $\sigma_t>0$ and $\sigma_\phi>0$. 
Then, in the gravity side, a wavepacket is created instead of the ``particle''.
and the response will also has finite width. 
(Although the particle-limit is given by  $\sigma_t\to 0$ and $\sigma_\phi \to 0$ while keeping $\Omega\gg 1/\sigma_t$ and $M\gg 1/\sigma_\phi$,  
we will take modest values for $\sigma_t$ and $\sigma_\phi$ because of numerical limitations.)
Since the subleading term of the asymptotic expansion of the bulk probe field corresponds to the response to $\mathcal{J}$ in the CFT, it is zero while the null geodesic is inside the bulk. However, it suddenly stands up just at the time the geodesic reaches the boundary. Thus, under the source~(\ref{src}), we can expect sharp signals of the response to be observed at discrete points, as shown in Fig.~\ref{nullg}(b).

\section{Transverse-field Ising model}
We consider the Ising model in transverse magnetic field on the $L$-site ring: 
\begin{equation}
 H=-J \sum_{j=1}^L \sigma_{j}^z \sigma_{j+1}^z-h\sum_{j=1}^L \sigma^x_{j}\ ,
\label{H0}
\end{equation}
where $\sigma_{j}^a$ $(a=x,y,z)$ is the Pauli matrix which acts on the $j$-th spin ($\sigma_{L+1}=\sigma_1$). 
This Hamiltonian is explicitly diagonalizable as summarized in appendix.\ref{TFI}. (See also Refs.\cite{Mbeng,Molignini} for nice reviews.)
The one-dimensional spin-1/2 chain can be mapped onto a fermionic system by the Jordan-Wigner transformation
\begin{equation}
 c_{j}=\frac{1}{2}\prod_{l=1}^{j-1}\sigma_{l}^x (-\sigma^z_{j}+i\sigma^y_{j})\ .
\label{ctosigma}
\end{equation}
Under the transformation Eq.~(\ref{ctosigma}), the Hamiltonian~(\ref{H0}) is rewritten as 
\begin{multline}
 H=-J\sum_{{j}=1}^{L}(c_{j}^\dagger c_{{j}+1}+c_{{j}+1}^\dagger c_{j} + c_{j}^\dagger c_{{j}+1}^\dagger + c_{{j}+1}c_{j})\\
-h\sum_{{j}=1}^L(1-2c_{j}^\dagger c_{j})\quad (\textrm{with } c_{L+1}=-c_1)\ ,
\label{Hcs}
\end{multline}
where we assumed that the total number of fermions is even $N=\sum_{j=1}^L c_j^\dagger c_j\in 2\bm{Z}$ since the ground state is always in the even-$N$ sector~\cite{Mbeng}.

Before showing the behavior of spacetime-localized response, let us see the QFT description of the lattice model~(\ref{Hcs}) in the continuum limit. To this end, we introduce the fermion field $\Psi(x_j)=c_j/\sqrt{a}$  
where $a$ is the lattice spacing. In the continuum limit $a\to 0$ while keeping the total length of the ring, $\ell=L a$, finite, 
the Hamiltonian becomes
\begin{equation}
 H=-\int^\ell_0 dx \left[\frac{v}{2}(\Psi^\dagger \frac{d}{dx}\Psi^\dagger-\Psi\frac{d}{dx}\Psi)+\delta\Psi^\dagger\Psi\right]\ ,
\label{Hconti}
\end{equation}
with $v=2Ja$ and $\delta=2(J-h)$.
This is just a field theory for the free Majorana fermion with the mass $\delta/v^2$. In the critical case $J=h$, the above Hamiltonian describes the CFT with central charge $c=1/2$. (See appendix.\ref{CFTising}.) Therefore, one can anticipate the occurrence of spacetime-localized response in the spin model~(\ref{H0})
by the virtue of FIG.\ \ref{theories},
particularly for a sufficiently large size $L$ and when $J\approx h$, with the caveat that the CFT is not strongly
coupled (as will be mentioned later). 

\begin{figure*}[t]
\centering
\includegraphics[scale=0.55]{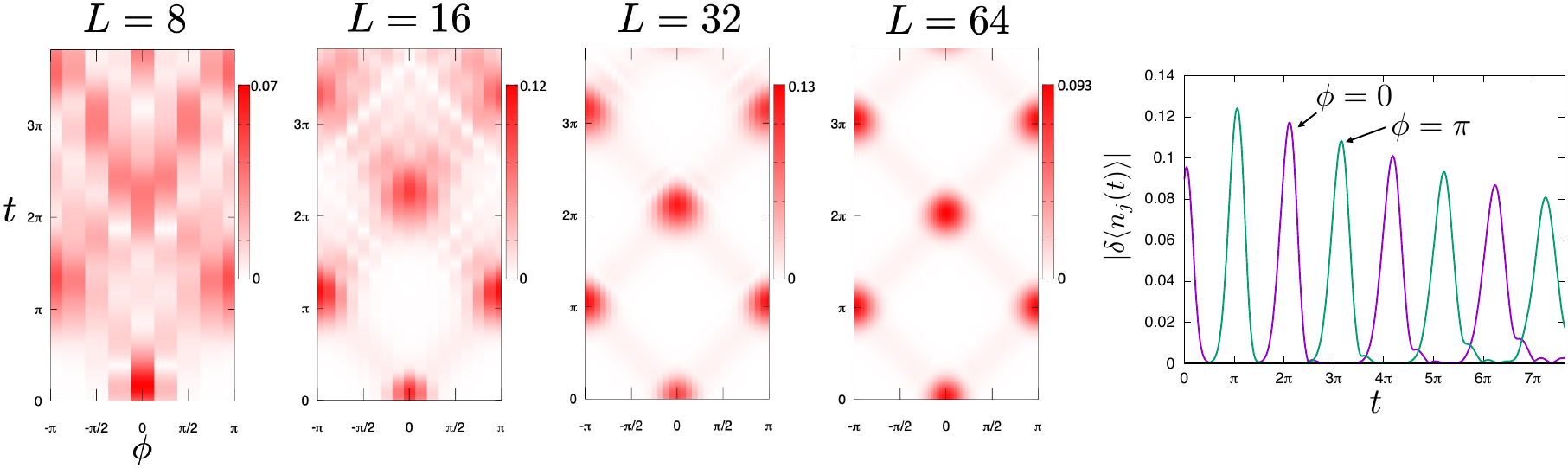}
\caption{Response for parameters $\delta=0$, $\sigma_t=0.4$, $\sigma_\phi=0.4$, $\Omega=5$ and $M=0$. The number of sites is varied as $L=8,16,32,64$. The color bar corresponds to $|\delta\langle n_j \rangle|$.
The rightmost panel shows time-dependence of the response for fixed $\phi$-slices at $\phi=0$ ($j={L/2}$) and  $\phi=\pi$ ($j={L}$). The number of sites is fixed as $L=32$.
}
\label{Lchange}
\end{figure*}

\section{Linear response theory}
Let us consider the linear response of the transverse-field Ising model with a finite $L$, whose Hamiltonian is given by Eq.~(\ref{H0}) or (\ref{Hcs}).
The perturbation of the Hamiltonian is
\begin{equation}
 \delta H (t)= -\sum_{{j}=1}^L \mathcal{J}_{{j}}(t)n_{j}\ .
\label{dH}
\end{equation}
where $n_j = c_j^\dagger c_j$ is the number operator of the fermion at the $j$-th site and $\mathcal{J}_{{j}}(t)=\mathcal{J}(t,\phi_{j})$ is the source function of the form Eq.~(\ref{src}).  Here, we introduced the spacial coordinate of the ${j}$-th site on the ring as 
\begin{equation}
 \phi_{j}=\frac{2\pi}{L}\left({j}-\frac{L}{2}\right)\ .
\end{equation}
From Eq.~(\ref{ctosigma}), the number operator $n_j$ is written as {$n_j=(1-\sigma^x_j)/2$}.
Thus, $\mathcal{J}_{{j}}(t)$ is regarded as the transverse magnetic field which depends on the time and space. 

The linear response of the ground state to the dynamic perturbation $\delta H (t)$ is given by
\begin{equation}
 \delta \langle  n_j(t)\rangle = -\sum_{l=1}^L \int_{-\infty}^\infty dt'  G_R(t-t',j-l) \mathcal{J}_l(t') \ ,
\label{linres}
\end{equation}
where $G_R(t-t',j-l)=-i \theta(t-t')\langle \, [n_j(t), n_l (t')]\, \rangle$ is the retarded propagator with $n_j(t)=e^{iHt} n_j e^{-iHt}$, and $\langle \cdots \rangle$ represents the expectation value with respect to the ground state.
We have an explicit expression for the retarded propagator as 
\begin{multline}
G_R(t,j)=\frac{2}{L^2}\theta (t) \sum_{k,k'\in K} u_{k} v_{k'} (u_k v_{k'}+u_{k'}v_{k})\\
\times \sin\big[(\epsilon_k+\epsilon_{k'})t-(k-k')j\big]\ ,
\label{GRexpression}
\end{multline}
where $\epsilon_k = 2J\{(\cos k-h/J)^2+\sin^2 k\}^{1/2}$ is the energy of the ``single particle state''. We have also defined 
$(u_{k},v_k)=\mathcal{N}(\epsilon_k+z_k,iy_k)$ with $z_k = 2(h-J\cos k)$ and $y_k = 2J\sin k$ where $\mathcal{N}$ is the normalization constant to make $|u_{k}|^2+|v_k|^2=1$. 
See appendix.\ref{RP} for the derivation of the above expression.
Using Eqs.(\ref{src}) and (\ref{GRexpression}) in Eq.~(\ref{linres}), we can compute the linear response. 

Note that, in the linear response theory, we simultaneously describe the formulations for the cosine (real) and sine (imaginary) parts of the source field in terms of the ``complex'' form of $\mathcal{J}_{l}(t^\prime)$. In the experiments, one has to measure the responses $\delta\langle  n_j(t)\rangle$ against $\Re [\mathcal{J}_l]$ and $\Im [\mathcal{J}_l]$ separately, and then combined the results into the form of $|\delta\langle  n_j(t)\rangle|$. 

\section{Spacetime-localized response}
Taking units of $v=1$ and $\ell=2\pi$, we have 
\begin{equation}
 a=\frac{2\pi}{L}\ ,\quad J=\frac{L}{4\pi}\ ,\quad h=J-\frac{\delta}{2}\ .
\end{equation}
The free parameters of the Hamiltonian are now given by $L$ and $\delta$.  
We take the amplitude of the source as $A=J\sqrt{2/(\sigma_t \sigma_\phi L)}$ so that 
$\sum_j \int dt |\mathcal{J}_{j}(t)|^2 \simeq J^2$ is satisfied.

Figure \ref{Lchange} shows the response $|\delta \langle  n_j(t)\rangle|$ for $\delta=0$, $\sigma_t=0.4$, $\sigma_\phi=0.4$, $\Omega=5$ and $M=0$. The number of sites are varied as $L=8,16,32,64$.
As shown in the figures, the response suddenly stands up around at the points indicated in Fig.~\ref{nullg}(b). 
(Since the source~(\ref{src}) has finite width, the response is not completely localized.)
This result shows that the spacetime-localized response can indeed be observed in the realistic spin model on a finite-size lattice.
Note that the behavior of the spacetime-localized response is already seen even for a small $L$ (around for $L\sim 16$). 
In appendix.\ref{TEQS}, we give a direct calculation of the quantum state after the perturbation~(\ref{dH}). 
We find that, by the time translation $t\to t+\pi$, 
the wavefunction is mapped onto the antipodes on $S^1$ with a partial sign-flip in the momentum space near the critical point. 
It follows that, after the time translation $t\to t+2\pi$, the same quantum state reappears.

We show the time dependence of the response for fixed $\phi_{j}=0,\pi$ in the rightmost panel of Fig.~\ref{Lchange}, in which the spacetime-localized response is clearly seen. 
According to the geodesic motion in the AdS$_3$, the peaks of the response should be at $t=0, 2\pi, 4\pi,\cdots$ for $\phi=0$ and $t=\pi, 3\pi, 5\pi \cdots$ for $\phi=\pi$, but they seem to appear a little later. The shift of the peak positions is caused by the finite-$L$ effect, and it actually gets smaller as $L$ increases.

In appendix.\ref{DetailLP}, we also summarize results of the linear response for $\delta\neq 0$ and $M\neq 0$.

In this paper, we have taken the unit of $v=\ell/(2\pi)=\hbar=1$. We can easily restore the dimensions of the quantities as 
$(t,\Omega,\sigma_t,\delta)\to (t/T,\Omega T,\sigma_t/T, T \delta /\hbar)$ 
where $T=\hbar L/(4\pi J)$. 
Note that $M$ and $\sigma_\phi$ are dimensionless quantities as they are. For example, in Fig.~\ref{Lchange}, we set $\Omega=5$. This implies $\Omega=5\times 4\pi J/(\hbar L)$. 

\section{On experimental realization}
The experimental realization and detection of our theoretical findings are feasible using Rydberg atoms trapped in an optical tweezers array~\cite{Browaeys:2020,Bernien:2017,Nogrette:2014,Ebadi:2021,Scholl:2021}. The state-of-the-art techniques developed in recent years have enabled us to simulate a programmable Ising-type quantum spin model with tunable interactions~\cite{Bernien:2017}, system sizes of up to hundreds~\cite{Ebadi:2021,Scholl:2021}, and arbitrary lattice geometries~\cite{Nogrette:2014,Scholl:2021}. The spacetime-localized response can be tested in a ring-shaped lattice~\cite{Nogrette:2014} of atoms near the quantum critical regime ($J\approx h$), achieved by global laser light that introduces the coupling between the ground and Rydberg states. The source fields, $\Re [\mathcal{J}_l]$ and $\Im [\mathcal{J}_l]$, could be implemented by temporary focused lasers with programmatically adjusted intensity for each spin on the ring within a range of $\sim \sigma_\phi$, or could be potentially manipulated more effectively with the help of a spatial light modulator~\cite{Nogrette:2014,Bijnen2015}. The response in $\langle \sigma_{j}^x(t)\rangle=1-2\langle n_{j}(t)\rangle$ of the individual atoms can be monitored at each time slice via the fluorescence imaging after inserting a global $\pi/2$ pulse. Additionally, systems of superconducting qubits~\cite{Morvan2022} and trapped ions~\cite{Jurcevic2014} should also offer another promising platform for realizing our theoretical proposal. 

\section{Conclusion and future directions}
In this paper, we introduced the concept of ``spacetime-localized response,'' a phenomenon where a particle appears to be transferred, travelling in the dual higher-dimensional geometry, and showed that this phenomenon actually takes place in the transverse-field Ising model on a lattice ring. This phenomenon is expected to be ubiquitous and working as a probe of the dual higher-dimensional geometry.

Although spacetime-localized response can be clearly understood in the gravity side, its physical interpretation in the spin system is not trivial. This would result from the long range correlation of the ground state of the spin system near the critical point. Revealing the physical interpretation of spacetime-localized response is an interesting future challenge.

The transverse-field Ising model reduces to the free QFT in the continuum limit as in Eq.~(\ref{Hconti}). 
There has been attempts for dual descriptions of free quantum theories. 
One of them is to consider a higher spin gravity equivalent to a free $N$-Majorana fermion representation~\cite{Ahn:2011pv,Gaberdiel:2011nt}.
Therefore, at least some higher spin gravity models may be a target for the experimental probe of the dual higher-dimensional spacetime.

Studying finite temperature effects is one of the most important future directions.
For a free QFT such as Eq.~(\ref{Hconti}), its finite temperature effects is trivial.
On the other hand, for the $SU(N)$ Heisenberg model for example, 
the continuum limit is a Wess-Zumino-Witten model~\cite{Affleck:1986bv} and can give non-trivial thermal effects. 
It would allows us to probe quantum black hole spacetimes through tabletop experiments.

Finally, let us suggest possible directions for the application of the spacetime-localized response.
1) It may provide a new method in spintronics  or magnonics. 
The spacetime-localized response might be used for carrying spin-wave packets from place to place, bypassing undesired operating elements existing on the way. Topological materials whose edge states are gapless CFT would be 
a suitable test ground for it. 
2)
It is tempting to suggest a similarity between the spacetime-localized response and a time crystal. Figure \ref{Lchange}
evidently forms a spacetime crystal-like structure in the two-dimensional spacetime. 
Although the spacetime-localized response is not a spontaneous breaking of time translation, the crystaline pattern formation would provide some novel holographic understanding of critical materials.
3)
Once at a finite temperature a holographic quantum black hole is realized, the quantum matter ring would serve as a quantum ``trash can'', when this ring is connected to quantum circuits. Black holes are the fastest scramblers \cite{Sekino:2008he}, and information is effectively lost, which could be efficiently used in quantum information science. These are just a list of interesting suggestions, and we like to explore them in the forthcoming papers.

%%%%%%%%%%%%%%%%%%%%%%%%%%%%%%%%%%%%%%%%%%%%%%%%%%%%%%%%%%%%%%%%%%%%%%%%%%%%%%%%%%%%%%%%%%%%%%%%%%%%%%%%%%%%%%%
\begin{acknowledgments}
We would like to thank Yasuaki Hikida, Akio Hosoya, Seiji Terashima and Takafumi Tomita 
for useful discussions and comments.
The work of M.\ B.\ was supported in part by JSPS Grant No.\ JPJSJRP20221202.
The work of K.\ M.\ was supported in part by JSPS KAKENHI Grant Nos.\ JP20K03976, JP21H05186 and JP22H01217.
The work of K.\ H.\ was supported in part by JSPS KAKENHI Grant Nos.\ JP22H01217, JP22H05111 and JP22H05115.
The work of D.\ T.\ was supported by Grant-in-Aid for JSPS Fellows No.\ 22KJ1944. 
The work of D.\ Y.\ was supported by JSPS KAKENHI Grants No. 21H05185, No. 22H01171, No. 23H01133, and No. 24K06890, and JST PRESTO Grant No. JPMJPR2118.
\end{acknowledgments}

\appendix

\section{Transverse-field Ising model}
\label{TFI}

The transverse-field Ising mode is a solvable spin model, which reduces to a conformal field theory (CFT) at the critical point.
In appendices, we give a brief review of the diagonalization of the Hamiltonian of the transverse-field Ising model. 
Analytical calculations of the two-point function and linear response are demonstrated. 
We present some supplementary results of the linear response which are not covered in the main text.
The quantum state after the perturbation of the Hamiltonian is also explicitly computed.
We take the CFT limit of the two point function and find that it coincides with that from the the general argument of the CFT.

The Hamiltonian of the transverse-field Ising model on the $L$-site ring is
\begin{equation}
 H=-J \sum_{i=1}^L \sigma_i^z \sigma_{i+1}^z-h\sum_{i=1}^L \sigma^x_i\ ,
\label{H0a}
\end{equation}
where $\sigma_i^a$ $(a=x,y,z)$ is the Pauli matrix which acts on the $i$-th spin and $\sigma_{L+1}=\sigma_1$.
In this section, we give a brief review of the diagonalization of the transverse-field Ising model. 
(See also Refs.~\cite{Mbeng,Molignini} for nice reviews.)
The one-dimensional spin-1/2 chain can be mapped onto the fermion system by the Jordan-Wigner transformation
\begin{equation}
 \sigma_i^x = 1-2c_i^\dagger c_i\ ,\quad 
-\sigma_i^z+i\sigma_i^y=2\prod_{j=1}^{i-1}(1-2c_j^\dagger c_j)c_i\ .
\label{JW}
\end{equation}
Its inverse transformation is given by 
\begin{equation}
 c_i=\frac{1}{2}\prod_{j=1}^{i-1}\sigma_j^x (-\sigma^z_i+i\sigma^y_i)\ .
\label{ctosigmaa}
\end{equation}
The operator $c_i$ satisfies the canonical anti-commutation relation, $\{c_i,c_j^\dagger\}=\delta_{ij}$.
By the Jordan-Wigner transformation, 
the transverse field Ising model reduces to the system of the fermions as
\begin{multline}
 H=-J\sum_{i=1}^{L}(c_i^\dagger c_{i+1}+c_{i+1}^\dagger c_i + c_i^\dagger c_{i+1}^\dagger + c_{i+1}c_i)\\
-h\sum_{i=1}^L(1-2c_i^\dagger c_i)\ ,
\label{Hcsa}
\end{multline}
where $c^\dagger_i$ and $c_i$ are the creation and annihilation operators of fermions at $i$-th site. %They satisfy $\{c_i,c_j^\dagger\}=\delta_{ij}$. 
We assume that the total number of fermions is even: $N=\sum_{j=1}^L c_j^\dagger c_j\in 2\bm{Z}$. Then, $c_i$ satisfies the anti-periodic boundary condition $c_{L+1}=-c_1$.

We apply the Fourier transformation of the operator $c_j$ as 
\begin{equation}
 c_j = \frac{1}{\sqrt{L}}\sum_{k\in K} e^{ikj} c_k\ .
\end{equation}
From the anti-periodic boundary condition in Eq.~(\ref{Hcsa}), the domain of the wave number $k$ is given by
\begin{equation}
 K=\left\{\frac{2\pi}{L} \left(n-\frac{1}{2}\right)  \bigg| n=-\frac{L}{2}+1,\cdots,\frac{L}{2}\right\}\ .
\label{Kdef}
\end{equation}
In the momentum space, the Hamiltonian becomes
\begin{multline}
 H=-J\sum_{k\in K} (2\cos k\, c_k^\dagger c_k + e^{ik} c_k^\dagger c_{-k}^\dagger + e^{-ik} c_{-k} c_k)\\
+ h \sum_{k\in K}(2 c_k^\dagger c_k-1)
\end{multline}
Rewriting the Hamiltonian~(\ref{Hcsa}) in terms of $c_k$, we find the coupling between modes with $k$ and $-k$. However, after the Bogoliubov transformation
\begin{equation}
 \mtx{c_k \\ c_{-k}^\dagger}=\mtx{u_k & -v_k^\ast \\  v_k  & u_k^\ast} \mtx{\gamma_k \\ \gamma_{-k}^\dagger}\ .
\label{defgamma}
\end{equation}
we obtain the diagonalized Hamiltonian as 
\begin{equation}
 H=E_0+\sum_{k\in K}\epsilon_k \gamma_k^\dagger \gamma_k \ .
\label{Hdiag}
\end{equation}
where 
\begin{equation}
\begin{split}
 &\epsilon_k = 2J\sqrt{(\cos k-\frac{h}{J})^2+\sin^2 k}\ ,\\
&\mtx{u_k\\v_k} = \frac{1}{\sqrt{2\epsilon_k (\epsilon_k+z_k)}}\mtx{\epsilon_k+z_k\\iy_k}\ ,
\end{split}
\end{equation}
with $z_k = 2(h-J\cos k)$ and $y_k = 2J\sin k$.
The constant term in Eq.~(\ref{Hdiag}) is the energy of the ground state given by $E_0=-\sum_{k\in K} \epsilon_k/2$.

From Eq.~(\ref{Hdiag}), we find that the ground state $|0\rangle$ is the state which is annihilated by $\gamma_k$: 
\begin{equation}
 \gamma_k |0\rangle = 0 \quad  (\forall k \in K)\ .
\label{goundstatedef}
\end{equation}
Excited states are constructed by multiplying the creation operators to the ground state as
\begin{equation}
 |\vec{m}\rangle=\prod_{k\in K} (\gamma_k^\dagger)^{m_k}|0\rangle\qquad (m_k=0\textrm{ or }1)\ ,
\end{equation}
where $\vec{m}$ represents the list of $m_k$. As assumed in Eq.~(\ref{Hcsa}), the total fermion number $\sum_{k\in K}m_k$ should be even. Their energy eigenvalues are
$E(\vec{m})=E_0+\sum_{k\in K} m_k \epsilon_k$.

\section{Retarded propagator and linear response}
\label{RP}

We define the two-point function in the transverse-field Ising model as
\begin{equation}
 C_{s}(t)=\langle n_{j+s}(t) n_j(0) \rangle \ .
\label{2ptdef}
\end{equation}
where $\langle \cdots \rangle $ is the expectation value with respect to the ground state. We denoted the number operator of the fermion at the $j$-th site  by $n_j =c_j^\dagger c_j$, and introduced its Heisenberg picture as $n_j(t)=e^{iHt}n_je^{-iHt}$. 
We can also express $n_j$ in terms of  $\gamma_k$ defined in Eq.~(\ref{defgamma}) as
\begin{multline}
n_j=\frac{1}{L}\sum_{k,k'\in K} 
e^{i(k'-k)j} \\
\times (u_k^\ast \gamma_k^\dagger -v_k \gamma_{-k}) (u_{k'} \gamma_{k'} -v_{k'}^\ast \gamma_{-k'}^\dagger)\ .
\label{nj_gamma}
\end{multline}
The two point function is rewritten as $C_s(t)=\langle0|  n_{j+s} e^{-i(H-E_0)t} n_j|0 \rangle$.
From Eq.~(\ref{nj_gamma}), we have
\begin{multline}
n_j|0\rangle
=-\frac{1}{L}\sum_{k,k'\in K}e^{i(k'-k)j}u_k^\ast v_{k'}^\ast |k,-k'\rangle\\
 + \frac{1}{L}  \sum_{k\in K} |v_k|^2 |0\rangle  \ ,
\label{n0}
\end{multline}
where we defined the two-particle state 
\begin{equation}
 |k,-k'\rangle \equiv  \gamma_k^\dagger \gamma_{-k'}^\dagger |0 \rangle\ .
\end{equation}
This satisfies 
\begin{equation}
\begin{split}
 &\langle0 |k,-k'\rangle=0\ ,\\
 &\langle p,-p'|k,-k'\rangle =\delta_{pk}\delta_{p'k'}-\delta_{p,-k'}\delta_{p',-k}\ .
\end{split}
\end{equation}
Since the ground state $|0\rangle $ and the two-particle state $|k,-k'\rangle$ are energy eigenstates, we also  have 
$e^{-i(H-E_0)t}|0\rangle = |0\rangle$ and $e^{-i(H-E_0)t}|k,-k'\rangle = e^{-i(\epsilon_k+\epsilon_{k'})t}|k,-k'\rangle$. 
From these relations, the two-point function is computed as
\begin{multline}
 C_{s}(t)=\frac{1}{L^2}\left(\sum_{k\in K} |v_k|^2\right)^2\\
-\frac{1}{L^2}\sum_{k,k'\in K}e^{-i(\epsilon_k+\epsilon_{k'})t+i(k-k')s}\\
\times u_{k} v_{k'} (u_k v_{k'}+u_{k'}v_{k})\ .
\label{2ptfunc}
\end{multline}

The regarded propagator is defined as 
\begin{equation}
 G_R(t-t',j-l)=-i \theta(t-t')\langle \, [n_j(t), n_l (t')]\, \rangle\ ,
\end{equation}
which is computed from the two point function~(\ref{2ptdef}) as $G_R(t-t',j-l)= -i\theta(t-t')(C_{j-l}(t-t')-C_{l-j}(t'-t))$. Thus, we obtain the explicit expression for the regarded propagator as
\begin{multline}
 G_R(t-t',j-l)=-\frac{i}{L^2}\theta (t-t')\\
\times \sum_{k,k'\in K} u_{k} v_{k'} (u_k v_{k'}+u_{k'}v_{k})\\
\times \sum_{r=\pm 1} r e^{ir(\epsilon_k+\epsilon_{k'})(t-t')-ir(k-k')(j-l)}\ .
\label{GRexpressiona}
\end{multline}
Under the perturbation of the Hamiltonian, 
\begin{equation}
 \delta H = -\sum_{l=1}^L \mathcal{J}_{l}(t)n_l\ ,
\label{deltaH}
\end{equation}
its linear response is given by
\begin{equation}
 \delta \langle  n_j(t)\rangle = -\sum_{l=1}^L \int_{-\infty}^\infty dt'  G_R(t-t',j-l) J_l(t') \ .
\label{linresa}
\end{equation}
We assume the time dependence of the source as 
\begin{equation}
 \mathcal{J}_l(t)=A_l\exp\left[
-\frac{t^2}{2\sigma_t^2}-i\Omega t
\right]\ .
\label{srca}
\end{equation}
Although above expression is complex, the real or imaginary part shall be taken to be implicit.
This is localized in time and oscillates with the frequency $\Omega$. 
$A_l$ describes the spacial dependence of the source.
Then, we can perform the $t'$-integration analytically in Eq.~(\ref{linresa}). 
Using the error function: $\textrm{erfc}(z)=2\pi^{-1/2}\int^\infty_z e^{-t^2}dt$, we can write the analytical form of the linear response as
\begin{multline}
 \delta \langle n_j(t)\rangle = \frac{i\sigma_t}{L^2}\sqrt{\frac{\pi}{2}} 
\sum_{r=\pm1,k,k'}ru_{k} v_{k'} (u_k v_{k'}+u_{k'}v_{k})\\
\times \left(\sum_{l=1}^L A_l e^{ir(k-k')l}\right)e^{-ir(k-k')j}\\
\times \exp\left[
-\frac{\sigma_t^2}{2}\{\Omega+r(\epsilon_k+\epsilon_{k'})\}^2+ir(\epsilon_k+\epsilon_{k'})t
\right]\\
\times \textrm{erfc}\left[-\frac{t}{\sqrt{2}\sigma_t}-\frac{i\sigma_t}{\sqrt{2}}\{\Omega+r(\epsilon_k+\epsilon_{k'})\}\right]\ ,
\label{resexp}
\end{multline}
For a given space dependence of the source $A_l$, we can numerically compute the summation of $k,k',r$ and $l$.

As the explicit expression for the spacial part of the source~(\ref{srca}), we use 
\begin{equation}
 A_l=A\exp\left[-\frac{\phi_l^2}{2\sigma_\phi^2} +iM\phi_l
\right]\ ,
\end{equation}
where $\phi_l=(2\pi/L)(j-L/2)$. 
Parameters $\Omega$ and $M$ correspond to the energy and orbital angular momentum of the created particle in the gravity side.

\section{Detailed analysis  of the linear Response}
\label{DetailLP}

\begin{figure}
\centering
\includegraphics[scale=0.4]{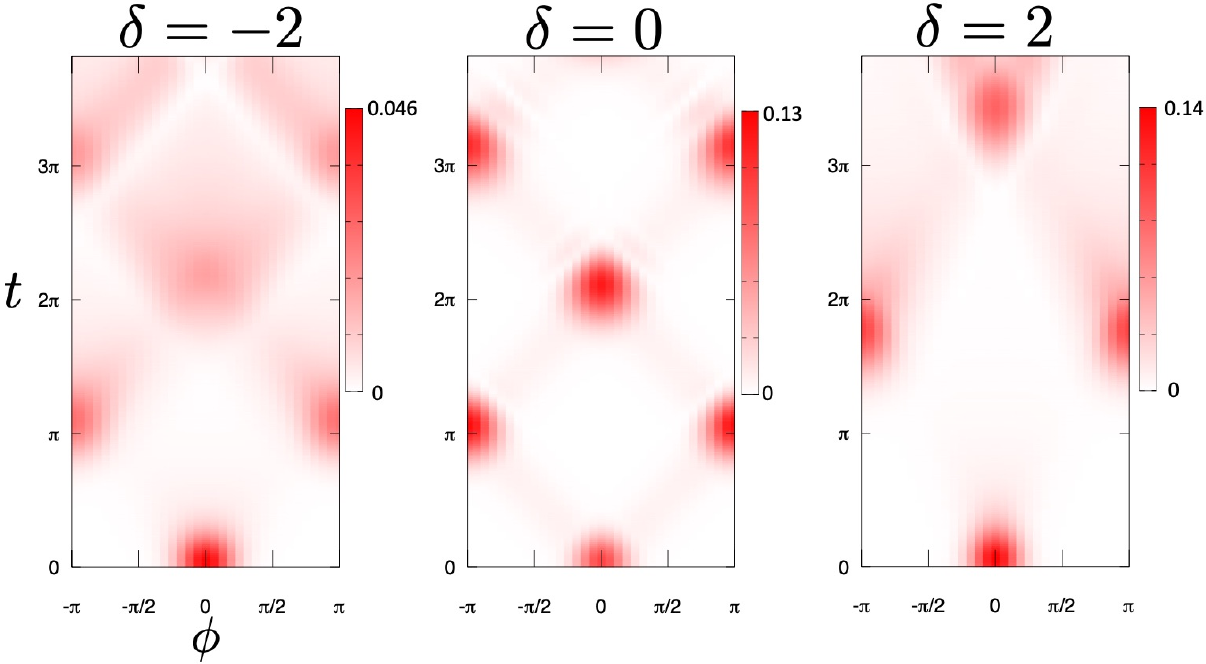}
\caption{Response for parameters $L=32$, $\sigma_t=0.4$, $\sigma_\phi=0.4$, $\Omega=5$ and $M=0$. $\delta=-2,0,2$. 
}
\label{Deltachange}
\end{figure}

In the main text, we focused on the response for the crital case $J=h$ ($\delta=0$) and for the zero orbital angular momentum $M=0$.
Here, the response in a wider parameter space will be studied. 
Figure \ref{Deltachange} shows the response for $\delta=-2,0,2$. Other parameters are set as 
$L=32$, $\sigma_t=0.4$, $\sigma_\phi=0.4$, $\Omega=5$ and $M=0$. %Now the mass of the Majorana fermion $\delta$ is changed as $\delta=-2,0,2$. 
Although the continuum limit of the system is not a CFT for $\delta\neq 0$, we can still see the behavior of the spacetime-localized response, although it gets blurred as $|\delta|$ increases. This is because the effect of the finite gap is negligible for $|\delta| \lesssim \Omega$ and the system would be approximated by the CFT.
We can also observe the shift of peak points of the response for $\delta\neq 0$.

\begin{figure}
\centering
\includegraphics[scale=0.4]{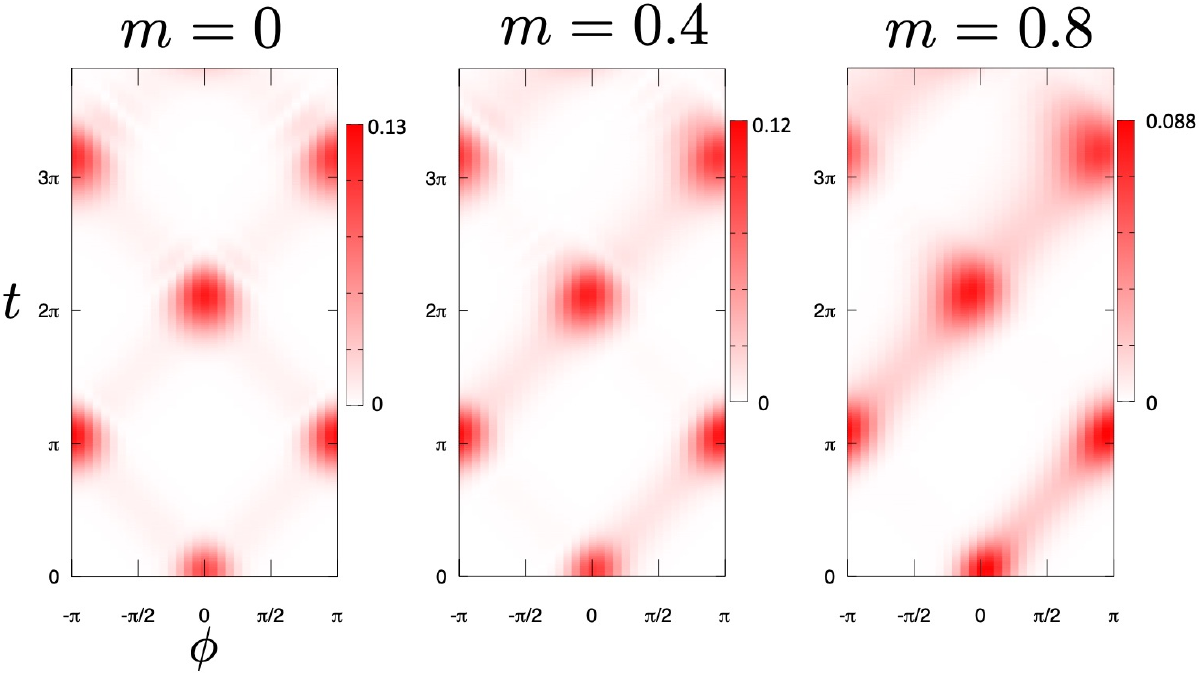}
\caption{Response for parameters 
$L=32$, $\delta=0$, $\sigma_t=0.4$, $\sigma_\phi=0.4$, $\Omega=5$ and $m=M/\Omega=0,0.2,0.4,0, 6,0.8$. 
}
\label{mchange}
\end{figure}

Figure \ref{mchange} shows the response for $L=32$, $\delta=0$, $\sigma_t=0.4$, $\sigma_\phi=0.4$, $\Omega=5$.  
The angular momentum of the null geodesic is varied as $m=M/\Omega=0,0.2,0.4,0.6,0.8$. 
The position of the peak points does not depend on $m$ so much. This is consistent with the result of null geodesics in AdS$_3$.
In the view of the gravity side, as the value of $|m|$ approaches 1, the null geodesic passes closer to the boundary. %(See Fig.~\ref{nullg}(a).)
Since the null geodesic is realized as a localized configuration of a probe field in AdS$_3$, it actually has a tail. When the particle is close to the AdS boundary, 
the response has a non-zero value because of the tail. This is the origin of the right moving tail in Fig.~\ref{mchange}.

\section{Conformal field theory limit of the transverse-field Ising model}
\label{CFTising}

Let us consider the the continuum limit of Eq.~(\ref{Hcsa}). 
We introduce the fermion field $\Psi(x_j)=c_j/\sqrt{a}$  where $a$ is the lattice spacing. For $a\to 0$, $\Psi(x)$ satisfies $\{\Psi(x),\Psi(x')\}=\delta(x-x')$. 
Then, the Hamiltonian is written as 
\begin{equation}
 H=-\int^\ell_0 dx \left[\frac{v}{2}(\Psi^\dagger \frac{d}{dx}\Psi^\dagger-\Psi\frac{d}{dx}\Psi)+\delta\Psi^\dagger\Psi\right]\ ,
\label{Hcontia}
\end{equation}
where 
\begin{equation}
 \ell=L a\ ,\quad v=2Ja=\frac{2J\ell}{L}\ ,\quad \delta=2(J-h)\ .
\label{contlim}
\end{equation}
This describe the theory for the free Majorana fermion with the mass $\delta/v^2$. 
For the critical case $J=h$, above Hamiltonian describes the conformal field theory (CFT) with central charge $c=1/2$.
Thus, 
the CFT limit of the transverse-field Ising model is given by
\begin{equation}
h= J\ ,\quad J\to \infty\ ,\quad  L\to\infty\ ,\quad a \to 0 \ ,
\label{CFTlim}
\end{equation}
with fixed $v$ and $\ell$ in Eq.~(\ref{contlim}). 
In the followings, we take the unit of $v=1$ and $\ell=2\pi$, i.e, $4\pi J/L=1$ and $La=2\pi$.

We consider the CFT limit of the two point function~(\ref{2ptfunc}).
Since the first term of Eq.~(\ref{2ptfunc}) is a constant, 
we will omit it in the following expressions. We can decompose the summation of $k$ and $k'$ as
\begin{equation}
 C_{s}(t)=-\frac{1}{L^2}\big(C_{s}^{uu}(t)C_{-s}^{vv}(t)
+C_{s}^{uv}(t)C_{-s}^{uv}(t)\big)\ ,
\label{2pt}
\end{equation}
where 
\begin{equation}
\begin{split}
 &C_s^{uu}(t)\equiv \sum_{k\in K} e^{-i\epsilon_kt+iks} u_k^2\ ,\\
&C_s^{uv}(t)\equiv \sum_{k\in K} e^{-i\epsilon_kt+iks} u_kv_k\ ,\\
&C_s^{vv}(t)\equiv \sum_{k\in K} e^{-i\epsilon_kt+iks} v_k^2\ .
\end{split}
\label{Csdef}
\end{equation}
In the critical case $J=h$, the dispersion relation becomes gapless as 
\begin{equation}
 \epsilon_k=4J\left|\sin \frac{k}{2}\right|\ .
\end{equation}
We also have
\begin{equation}
 u_k=\sqrt{\frac{1+|\sin\frac{k}{2}|}{2}}\ ,\quad
 v_k=\frac{i\cos\frac{k}{2}\, \textrm{sgn}(k)}{\sqrt{2(1+|\sin\frac{k}{2}|)}} \ .
\end{equation}

For the regularization, we shift the time coordinate to the complex plane as $t\to t-i\varepsilon$. 
For the shift parameter $\varepsilon$, 
we assume $1/J\ll \varepsilon$. Although this is bounded from below, after taking the CFT limit~(\ref{CFTlim}), we can eventually take the limit of $\varepsilon\to +0$.
Due to the imaginary part of the time coordinate, only the region of $|k|\lesssim 1/(J\varepsilon)\ll 1$ contributes in the summation of Eq.~(\ref{Csdef}). 
In this region, we can write $\epsilon_k\simeq 2J|k|$, $u_k\simeq 1/\sqrt{2}$ and $v_k\simeq i \,\textrm{sgn}(k) /\sqrt{2}$.
Thus, we have
\begin{equation}
\begin{split}
&C_s^{uu}(t-i\varepsilon)\to \frac{1}{2}\sum_{k\in K} e^{-2iJ |k| (t-i\varepsilon)+iks}\\
&= \frac{1}{2}\sum_{n=-\infty}^\infty e^{- i |n-1/2| (t-i\varepsilon)+ i (n-1/2) \phi}\\
&=\frac{1}{4i}\left(
\frac{1}{\sin \frac{(t-i\varepsilon)-\phi}{2}}+\frac{1}{\sin \frac{(t-i\varepsilon)+\phi}{2}}
\right)\ ,
\end{split}
\end{equation}
where we introduced the coordinate of the $s$-th spin site as $\phi=2\pi s/L$.
Similarly, we also obtain
\begin{equation}
\begin{split}
 C_s^{uv}(t-i\varepsilon)&\to \frac{1}{4}\left(
\frac{1}{\sin \frac{(t-i\varepsilon)-\phi}{2}}-\frac{1}{\sin \frac{(t-i\varepsilon)+\phi}{2}}
\right)\\
C_s^{vv}(t-i\varepsilon)&\to -\frac{1}{4i}\left(
\frac{1}{\sin \frac{(t-i\varepsilon)-\phi}{2}}+\frac{1}{\sin \frac{(t-i\varepsilon)+\phi}{2}}
\right)
\end{split}
\end{equation}
Therefore, from Eq.~(\ref{2pt}), the CFT limit of the two point function becomes
\begin{equation}
 C_s(t-i\varepsilon) \to -\frac{1}{4L^2}\frac{1}{\sin \frac{(t-i\varepsilon)+\phi}{2}\sin \frac{(t-i\varepsilon)-\phi}{2}}\ .
\label{2ptlim}
\end{equation}

Let us consider the two point function from the general argument of CFT. 
For Euclidean CFT in $\bm{R}^2$, the two-point function for the operator with conformal weight $(h,\bar{h})$ is given by
\begin{equation}
 \langle \mathcal{O}(z_1,\bar{z}_1) \mathcal{O}(z_2,\bar{z}_2)\rangle =\frac{1}{z_{12}^{2h}\bar{z}_{12}^{2\bar{h}}}\ ,
\label{OO}
\end{equation}
where $z_{12}=z_1-z_2$.
Here we will consider the spinless field: $h=\bar{h}=\Delta/2$. 
We can move to the CFT in the cylinder $\bm{R}\times S^1$ by the conformal transformation, $z=e^{-iw}$.
From $\mathcal{O}(z,\bar{z})=(\partial z/\partial w)^{-h}(\partial \bar{z}/\partial \bar{w})^{-\bar{h}}\mathcal{O}(w,\bar{w})$ and Eq.~(\ref{OO}), we obtain
\begin{equation}
\langle \mathcal{O}(\tau,\phi) \mathcal{O}(0)\rangle \propto \frac{1}{\left(\sin \frac{\phi+i\tau}{2} \sin\frac{\phi-i\tau}{2}\right)^\Delta}\ ,
\end{equation}
where we write the complex coordinate in $\bm{R}\times S^1$  as $w=\phi+i\tau$. 
The two-point function in the Lorentzian signature is given by the analytic continuation of the Euclidean time as $\tau=it+\varepsilon$~\cite{Hartman:2015lfa,Bautista:2019qxj}. 
Thus, we have
\begin{equation}
 \langle \mathcal{O}(t,\phi) \mathcal{O}(0)\rangle\propto
\frac{1}{\left(\sin \frac{(t-i\varepsilon)+\phi}{2}\sin \frac{(t-i\varepsilon)-\phi}{2}\right)^\Delta}\ .
\end{equation}
This coincides with Eq.~(\ref{2ptlim}) by setting $\Delta=1$.

\section{Time evolution of quantum state}
\label{TEQS}

By the perturbation of the Hamiltonian~(\ref{deltaH}), the quantum state is perturbed as $|\psi(t)\rangle = e^{-iE_0 t} (|0\rangle + |\delta\psi(t)\rangle)$ where we take the ground state as the non-perturbative state. Solving the Shr\"{o}dinger equation in the first order in the perturbation, we have
\begin{equation}
 i|\delta\psi(t)\rangle = e^{-i(H-E_0)t}\int^t_{-\infty}dt' e^{i(H-E_0)t'} \delta H(t') |0\rangle \ ,
\end{equation}
where we take the lower bound of the integration so that $|\delta\psi(t)\rangle\to 0$ for $t\to -\infty$.
As the source function, we take the real part of Eq.(\ref{srca}):
\begin{equation}
 J_l(t)=\frac{1}{2}\sum_{r=\pm 1} A_l^r e^{-ir\Omega t}e^{-\frac{t^2}{2\sigma_t^2}}\ ,
\label{Jlreal}
\end{equation}
where we introduced $A_l^{+1}=A_l$ and $A_l^{-1}=A_l^\ast$. 
We will consider the sufficiently late time after the stimulus: $t\gg \sigma_t$. Then, the perturbation of the state becomes
\begin{equation}
 i |\delta\psi(t)\rangle \simeq e^{-i(H-E_0)t}\int^\infty_{-\infty}dt' e^{i(H-E_0)t'} \delta H(t') |0\rangle \ .
\label{psiperturb}
\end{equation}
Then, from Eqs.(\ref{n0}), (\ref{deltaH}) and (\ref{Jlreal}), we can perform the Gaussian integration and obtain
\begin{equation}
 |\delta\psi(t)\rangle =\Psi_0 |0\rangle + \sum_{k,k'\in K} \Psi_{kk'}(t)|k,-k'\rangle \ ,
\label{dPsisol}
\end{equation}
where 
\begin{equation}
\begin{split}
&\Psi_{kk'}(t)=-i\frac{\sqrt{2\pi}\sigma_t}{2L}\sum_{r=\pm1}(\sum_l A_l^r e^{i(k'-k)l})\\
&\qquad\qquad \times e^{-\frac{\sigma_t^2}{2}(\epsilon_k+\epsilon_{k'}-r\Omega)^2-i(\epsilon_k+\epsilon_{k'})t}u_k^\ast v_{k'}^\ast\ ,\\
&\Psi_0=i\frac{\sqrt{2\pi}\sigma_t}{2L}(\sum_{r=\pm1,l} A_l^r )\sum_{k\in K}|v_k|^2 e^{-\frac{\sigma_t^2}{2}\Omega^2}\ .
\end{split}
\label{PQ}
\end{equation}
Note that $\Psi_{kk'}(t)$ and $\Psi_0$ can be regarded as wavefunctions in the momentum space.

In the main text, we found the spacetime-localized response to a special source function. 
Especially, the spacetime-pattern of the response seems invariant under the discrete spacetime translation: 
\begin{equation}
 t\to t+\pi\ ,\quad \phi\to \phi+\pi\ .
\label{disctrans}
\end{equation}
Is the quantum state itself be copied to the antipodes in $S^1$ after the time translation $t\to t+\pi$?
To see this, we apply the discrete spacetime translation to the perturbed quantum state $|\delta\psi(t)\rangle$. 
The time-translation operator is given by $U(T)=e^{-i(H-E_0)T}$. 
We also define the operator to translate the site by $d\in \bm{Z}$ as
\begin{equation}
 \mathcal{T}(d)|k_1,\cdots,k_n\rangle = e^{-i(k_1+\cdots+k_n)d}|k_1,\cdots,k_n\rangle\ ,
\end{equation}
where $|k_1,\cdots,k_n\rangle \equiv  \gamma_{k_1}^\dagger \cdots\gamma_{k_n}^\dagger |0 \rangle$. 
The point reflection on $S^1$, $\phi\to \phi+\pi$, is performed by the operation of $\mathcal{T}(L/2)$.
After the discrete spacetime translation~(\ref{disctrans}), the quantum state becomes
\begin{multline}
U(\pi)\mathcal{T}(L/2) |\delta\psi(t)\rangle\\=\Psi_0 |0\rangle
+
\sum_{k,k'\in K} \alpha_{kk'} \Psi_{kk'}(t) |k,-k'\rangle   \ .
\end{multline}
where 
\begin{equation}
 \alpha_{kk'}\equiv e^{-i(\epsilon_k+\epsilon_{k'})\pi-i(k'-k)L/2}
\label{alphadef}
\end{equation}
Thus, the wavefunction get the phase $\alpha_{kk'}$ after the discrete spacetime translation.

As in section.\ref{CFTising}, in the CFT limit, the energy spectrum is given by $\epsilon_k\simeq L|k|/(2\pi)$. Thus, in the exponent of Eq.(\ref{alphadef}), we have
\begin{equation}
\begin{split}
 &\epsilon_k+\epsilon_{k'}+(k'-k)\frac{L}{2\pi}\simeq (|k|-k + |k'|+k)'\frac{L}{2\pi}\\
&=\frac{L}{\pi} \times 
\begin{cases}
 k' & (k>0,k'>0)\\
 0 & (k>0,k'<0) \\
 -k+k' & (k<0,k'>0)\\
 -k & (k<0,k'<0)
\end{cases}
\ .
\end{split}
\end{equation}
Since $k$ and $k'$ are components of $K$ defined in Eq.(\ref{Kdef}), 
above expression becomes odd-integer for $kk'>0$ and even-integer for $kk'<0$. It follows that, in the CFT limit, the phase $\alpha_{kk'}$ is simply written as
\begin{equation}
 \alpha_{kk'}\simeq -\textrm{sgn}(kk')\ .
\label{alphaCFT}
\end{equation}
Therefore, by the discrete spacetime translation~(\ref{disctrans}), the quantum state is not invariant but gets the partial sign-flip in its wavefunction.

Why does the response appear invariant under discrete spacetime translation~(\ref{disctrans}), while the quantum state does not exhibit such invariance?
One can directly compute the linear response from the perturbed quantum state~(\ref{dPsisol}) as
\begin{equation}
 \delta\langle n_j(t)\rangle = 2\textrm{Re}[ \langle \delta \psi(t)|n_j|0\rangle ]\ .
\end{equation}
From Eqs.(\ref{n0}) and (\ref{dPsisol}), we obtain
\begin{multline}
 \langle \delta \psi(t)|n_j|0\rangle \\
= \frac{1}{L}\sum_{kk'}\Psi_{kk'}^\ast(t) e^{i(k'-k)j}(u_kv_{k'}+u_{k'}v_k)\\
+\frac{1}{L}  \Psi_0 \sum_{k\in K} |v_k|^2\ .
\end{multline}
The last term is pure imaginary and does not contribute to the response.
In the summand of above expression, there is $u_k v_{k'}(u_kv_{k'}+u_{k'}v_k)$.
In the CFT limit, we can write $u_k\simeq 1/\sqrt{2}$ and $v_k\simeq i \,\textrm{sgn}(k) /\sqrt{2}$. Thus, we have
\begin{equation}
 u_k v_{k'}(u_kv_{k'}+u_{k'}v_k)\simeq \frac{i}{4}(1+\textrm{sgn}(kk'))\ .
\label{uvuvCFT}
\end{equation}
By the discrete spacetime translation~(\ref{disctrans}), the wavefunction is changed as $\Psi_{kk'}\to \alpha_{kk'}\Psi_{kk'}$. 
From Eq.(\ref{alphaCFT}) and (\ref{uvuvCFT}), we obtain
\begin{equation}
\alpha_{kk'}u_k v_{k'}(u_kv_{k'}+u_{k'}v_k)\simeq -u_k v_{k'}(u_kv_{k'}+u_{k'}v_k)\ .
\end{equation}
It follows that the response just change the signature as $\delta\langle n_j(t)\rangle \to -\delta\langle n_j(t)\rangle$. Therefore, by the translation~(\ref{disctrans}), the response is copied to the antipodes in $S^1$ after sign-flip. 
\newpage

%%%%%%%%%%%%%%%%%%%%%%%%%%%%%%%%%%%%%%%%%%%%%%%%%%%%%%%%%%%%%%%%%%%%%%%%%%%%%%%%%%%%%%%%%%%%%%%%%%%%%%%%%%%%%%%
% The \nocite command causes all entries in a bibliography to be printed out
% whether or not they are actually referenced in the text. This is appropriate
% for the sample file to show the different styles of references, but authors
% most likely will not want to use it.
% \nocite{*}

\bibliography{bunken}% Produces the bibliography via BibTeX.

\end{document}